\documentclass[12pt]{iopart}

\usepackage{enumitem}

\usepackage[bottom]{foot misc}

\usepackage{graphicx}
\usepackage{amsthm}   
\usepackage{amssymb}  
\usepackage{mathrsfs} 
\usepackage{stmaryrd} 
\usepackage{txfonts} 

\usepackage{hyperref}  
\usepackage{cleveref}
\usepackage{tikz}
\usetikzlibrary{circuits.logic.US, positioning}

\newcommand{\BBB}{{\mathbb{B}}}

\newcommand{\ZZZ}{{\mathbb{Z}}}

\usepackage{color}  

\begin{document}

\title{What does it mean for a system to compute?}

\author{David H. Wolpert}

\address{Santa Fe Institute, Santa Fe, NM 87501, USA}
\address{Complexity Science Hub, Metternichgasse 8, 1030, Vienna, Austria}
\address{ICTP – The Abdus Salam International Centre for Theoretical Physics, Strada Costiera 11, 34151 Trieste, Italy}
\address{Albert Einstein Institute for Advanced Study, New York, NY, USA}

\ead{david.h.wolpert@gmail.com}

\author{Jan Korbel}

\address{Complexity Science Hub, Metternichgasse 8, 1030, Vienna, Austria}
\ead{korbel@csh.ac.at}

\begin{abstract}
Many real-world dynamic systems, both natural and artificial, are understood to be performing
computations. For artificial dynamic systems, explicitly designed to perform computation --- such as digital computers --- by construction, we can identify which aspects of the dynamic system match the input and output of the computation that it performs,  as well as the aspects of the dynamic system that match the intermediate logical variables of that computation.
In contrast, in many naturally occurring dynamical systems that we understand to be computers, even though we neither designed nor constructed them
 --- such as the human brain --- it is not \textit{a priori} clear 
how to identify the computation we presume to be encoded in the dynamic system. Regardless of their origin, dynamical systems capable of computation can, in principle, be mapped onto corresponding abstract computational machines that perform the same operations. 
In this paper, we begin by surveying a wide range of dynamic systems whose computational properties have been studied. We then introduce a very broadly applicable framework for identifying what computations(s) are
emulated by a given dynamic system. 
After an introduction, we summarize key examples of dynamical systems whose computational properties have been studied. 
We then introduce a very broadly applicable framework that defines the computation
performed by a given dynamical system in terms of maps between that system's evolution and the evolution of an abstract computational machine.
We illustrate this framework with several examples from the literature, in particular discussing why some
of those examples do \textit{not} fully fall within the remit of our framework. We also briefly discuss several
related issues, such as uncomputability in dynamical systems, and how to quantify the ``value of computation'' in naturally occurring computers. 
We conclude with a discussion of some of the promising directions for future research.
\end{abstract}

%
%
%
%
%

\section{Introduction}

\subsection{Background}
\label{sec:background}

From neural networks in simple organisms to modern cloud computer systems to human social systems, computational processes are pervasive in both natural and artificial dynamic systems~\cite{navlakha2011algorithms,haken2012computational,Chu2018computation,wolpert2024computational}. 
Centuries of research have significantly advanced our theoretical understanding of some of these computational processes.
In particular, mathematical developments have deepened our understanding of the computability~\cite{sipser1996introduction}, computational complexity~\cite{arora2009computational}, and (resource-bounded) Kolmogorov complexity~\cite{arora2009computational,li2008introduction} of computational tasks. These developments are exemplified by deep issues
like the famous \textbf{P} versus \textbf{NP} problem. 

In addition, there has been great progress in our ability to design 
and then physically construct dynamic systems to implement specific computations. To do this, we identify the initial (perhaps coarse-grained) state of 
the dynamic system with the desired input to the algorithm, and its final state (if it exists) is identified with the output of the computation. So changing
the actual dynamics of the system constitutes changing the computation itself.

Crucially, in such human-constructed computers, we choose the \textbf{decoding} map, taking the degrees of freedom of the dynamic system to the logical
variables in the abstract computer we wish to view that system
as implementing. This means that the relationship between such a system’s dynamics and the computer it is 
implementing is explicitly known before the dynamic system starts its evolution. In short,
we have an a priori \textit{computational blueprint} for mapping the dynamics of the dynamic system to that of a computer.

As a point of contrast, consider naturally occurring dynamic systems that are often viewed as performing computations, even 
though humans did not explicitly construct those systems to perform some specific computation. Examples of such dynamic systems range from individual cells to entire brains to turbulent flows to 
whole human societies. Even systems like off-equilibrium spin networks have been viewed as computers~\cite{meijers2021behavior}.

In these naturally occurring systems, there is no pre-specified {decoding} map taking the degrees of freedom of the dynamic system
into the logical variables of an abstract computational machine.
Moreover, the same dynamic system can often be viewed as performing different computations, depending on how we choose to 
map the dynamic system's variables to logical variables in a computational machine, i.e., how to decode those variables. 
(This is true of both human-constructed dynamic systems and naturally occurring ones.) More precisely, changing that map not only changes how we interpret the dynamic system's initial state as the input to a computer; it can also change how we interpret the subsequent dynamics 
of that system as a computation unfolding. 

In fact, many (fixed) dynamic systems can be viewed as performing an infinite set of
different computations, depending on our choice of the decoding map. A proof sketch of this is given by the following argument
(see also~\cite{piccinini2021computation}):
\begin{proof}
The claim is established essentially just by defining terms.

First, 
for any universal Turing machine $U$,
write $(p, x)$ for an arbitrary input to $U$, where $p$ is the specification of an arbitrary TM and $x$ is a
bit string in the halting set of $p$.
So we can write $U(p, x)$ for the output tape of $U$ when it reaches a halting state after starting with input $(p, x)$,
and $U(p, x) = p(x)$, the output of $p$ when it reaches a halting state after starting with input $x$.

It is well-known that there are many pairs of a naturally occurring dynamic system $\Xi$ and an associated UTM $U$ such that 
the state space $X$ of $\Xi$ 
contains two non-overlapping countably infinite subsets $A, B$ with the following properties:
\begin{enumerate}
\item There is a surjective map $\alpha_A$ whose domain is
the set of all possible $(p, x)$ in the halting set of $U$ and whose image is all of $A$;
\item There is a bijective map $\alpha_B$ whose domain is the set of all possible $U(p, x)$ 
and whose image is all of $B$;
\item For all pairs $(p, x)$, there exists at least one state $x_0 \in A$ such that the first state in $B$
that $\Xi$ reaches after being in state $\alpha_A(x_0)$ is $\alpha_B(U(p, x)) = \alpha_B(p(x))$. 
\end{enumerate}


Let $p'$ be some specific TM that we wish to be implemented by the dynamics of
$\Xi$ if we use some appropriate decoding map of the states of $\Xi$. Formally, this means
that there is some countably infinite subset $C \subset X$, some (possibly finite) $D \subset X$, and associated 
maps $\alpha_C$ and $\alpha_D$, such that for all  bit strings $x_0$ in the halting set
of the TM $p'$, if $\Xi$ is in state $\alpha_C(x_0)$ at some time
then the first state that $\Xi$ reaches in $D$ after then is $\alpha_D(p'(x))$.

The claim is immediate, simply by taking $C := \{(p, x_0) \in A : p = p'\}$, $\alpha_C(x) := \alpha_A(p', x)$, 
$D = \{y \in B : \exists x \; \alpha_B(U(p', x)) = y\}$, and for all $x \in D$, $\alpha_D(x) := \alpha_B(U(p', x))$.

Note, we make no assumption in the definition (i, ii, iii) above 
concerning the dynamics of $\Xi$ for any time
after it first reaches $B$. We also make no assumptions about the dynamics of $\Xi$ once it leaves $A$ but before it reaches $B$. So our definition does not concern the relationship between the step-by-step
evolution of $\Xi$ and the  step-by-step evolution of $p$.

In this, the definition in (i, ii, iii) above is weaker than the one considered below in~\cref{sec:formal_def}. 
However, with some more work, the same result that is proved here for the definition of ``computationally
universal dynamic system'' in (i, ii, iii) can be extended to apply to the definition  in~\cref{sec:formal_def}.
\end{proof}

\subsection{Constructed versus non-constructed computers}


Our goal with this paper is to formalize exactly how one can decode the states of a dynamic system into the logical variables in a computer, and then to review examples of such dynamic systems. To begin, note that there are two types of dynamic systems we can interpret as performing a computation.

In the first type, the decoding of some of the dynamic system's variables as the logical variables of the computational machine is known ahead of time. In this paper, we are interested in
such dynamic systems that are physical devices arising in the real world, where the decoding
is in fact explicitly used by a human engineer or scientist who has designed and then constructed that physical device,
with the explicit goal of using it to implement that computational machine.
We will refer to dynamic systems of this type as \textbf{constructed computers}. Ultimate examples of constructed computers are the digital devices that underlie modern society. See, e.g., \cite{cotler2024computationaldynamicalsystems} for a recent discussion.

The second type of dynamic system does not involve a human scientist designing 
and then constructing a physical device with an explicit goal of having it implement a desired computation. Rather, the scientist investigating such systems must infer or impose a decoding of some of the variables of that dynamic system as the logical variables of a computational machine.
We will refer to dynamic systems of this second type as \textbf{non-constructed computers}. 
For completeness, we stipulate that the human scientist analyzing a non-constructed computer has
no knowledge of any \textit{a priori}, privileged decoding of some of its physical variables as the logical variables of 
a computational machine\footnote{As an example of how this generality could be relevant, albeit a fanciful one, we could imagine that humanity discovers an alien computer, along with an instruction manual for using it. In this case, no human has constructed the dynamic system, but the ``correct'' encoding is known ahead of time to those humans. }.

Many (though not all) non-constructed computers occur in biological systems.
Much of the work in the literature on such computers involves the scientist choosing a way to decode its physical variables so that the resultant computational machine 
has some desired computational behavior, e.g., so that it is computationally universal. Quintessential examples
of such non-constructed computers are found throughout biology, including brains, eusocial insect colonies,
and individual cells.

This distinction between constructed and non-constructed computers is not as clear-cut as one might wish. 
In particular, suppose we modified the characteristics of a turbulent flow found in nature to have it implement
one computation rather than another. Strictly speaking, this is a constructed computer. However, because we start with a physical system already found in nature, there are aspects of this system
that overlap with those of non-constructed computers.

\section{Focus of this paper}
How should we identify the computation done by a dynamical system when it is non-constructed, and so we don’t have a pre-specified way of mapping its 
states to logical variables? 
Is there a principled way to identify what computation (or set of computations) is performed by an arbitrary (time-stationary) dynamic system without pre-specifying how we decode its states into logical variables? 

In general, we can identify many (often infinite) different
computations with the dynamics of any given non-constructed dynamic system~\cite{piccinini2021computation}, even if that dynamic system is not
computationally universal (so that the result derived in~\cref{sec:background} does not apply). How should we choose one of the many different computations that are consistent with the observed behavior of such a dynamic system, and privilege it as ``the'' computation that the system is performing~\cite{urai2022large}?

As ill-posed as this question is, answering it is a necessary first step to being able to construct a general framework for analyzing what computation(s) are performed by any dynamical system.

In this paper, we illustrate these issues by considering a broad range of non-constructed computers. This focus means that we do not consider nonstandard types of artificial (constructed) computers, like neuromorphic computers~\cite{furber2016large,schuman2022opportunities,markovic2020physics}, 
liquid brain computers, human-constructed
analog computers, etc. Just like standard computers, which are digital and typically use von Neumann architectures, the human designer of such non-standard artificial computers determines ahead of time how to identify the variables in the dynamic 
system with logical variables in an associated computation. This is true of all constructed computers, essentially by definition. Similarly, due to the limits of space, we do not explicitly consider quantum computation~\cite{deutsch1992rapid,feynman1986quantum,nielsen2006quantum}, nor analog computation (see~\cite{siegelmann1995computation,siegelmann1998analog} and references therein). We note, though, 
that the formalism we present in~\cref{sec:formal_def} is broad enough to encompass all of these. 

As a final comment, we emphasize that we do \textit{not} restrict ourselves to dynamic systems that are either Turing universal and/or that can exhibit uncomputable behavior. 
(Those specific issues are briefly discussed in~\cref{sec:TM} below.) Our interest is in dynamic
systems that can be interpreted as implementing any computational machine considered in computer science (CS) theory
even if it is weaker than universal Turing machines (TMs), e.g., any machine from the Chomsky hierarchy.  

The structure of the rest of the paper is as follows: we begin in~\cref{sec:examples} with a quick review 
of some of the non-constructed computers that have been investigated in the literature. As mentioned above, in general, many (often infinitely many) different computational machines are implemented by a given non-constructed computer. Accordingly, in~\cref{sec:formal_def} we introduce a broadly applicable framework for identifying the set of all possible computational machines that are implemented by any given non-constructed dynamic system.

In~\cref{sec:illustrations} we then
briefly illustrate our framework on some of the examples of non-constructed computers found in the literature. Next
in~\cref{sec:TM} we present a very brief summary of earlier work on TMs implemented by non-constructed computers. 
We end in~\cref{sec:future} with a discussion of the many instances of non-constructed computers that neither our framework nor any other framework we are aware of fully captures, and so might be quite fruitful topics for future research.

\section{Examples of dynamical systems that implement computation}
\label{sec:examples}

In this section, we discuss several examples of dynamic systems that can be seen as performing computation. The set of examples we present is meant to be illustrative rather than exhaustive; there are many other examples of dynamic systems that are capable of computation. We start with a summary
of a few non-constructed computers:

\begin{itemize}
\item \textbf{Neuroscience.} While it is common to compare a brain to a computer metaphorically, it has
also been suggested that a brain can be viewed as a computer in a more formal sense \cite {maley2022and}.
In particular, there has been extensive discussion of the hypotheses that the brain is operating near a critical point 
 \cite{Chialvo2010,hesse2014self,Shew15595,Brochini2016}, and
that it exhibits quasi-critical dynamics \cite{PhysRevLett.126.098101}.
This has led many researchers to apply information-theoretic concepts like predictive coding 
to analyze neuronal dynamics, see, e.g., \cite{clark2013whatever,doi:10.1073/pnas.1912340117,shain2020fmri,chalk2018toward}. Similarly, recent studies have focused on the computational power of the human brain \cite{gebicke2023computational}.
 
\item \textbf{Groups of multiple interacting biological organisms.}
There are many instances in biology of \textit{groups} of multiple biological organisms that are commonly viewed as performing computation.
One canonical example is eusocial insect colonies. The collective behavior of such colonies, induced by
processes like stigmergy~\cite{Khuong2016,HEYLIGHEN20164}, the ``waggle dance'' of bees~\cite{doi:10.1126/science.ade1702}, etc., is often interpreted as performing computation. Indeed, much research has been
designing computational algorithms to perform optimization explicitly inspired by stigmergy~\cite{Salman2024,Nichol_2024,theraulaz1999brief}.

There are many other similar examples. A notable one is
animal swarms, e.g., of fish, starlings, etc.~\cite{BALLERINI2008201}, whose joint dynamics has been viewed as performing a computation. These swarms coordinate primarily through visual means.
Slime molds, in contrast, are also often viewed as performing computation~\cite{Adamatzky03042015}, though visual processing is not involved. Another striking example of naturally occurring dynamic systems, which are often decoded as computing systems, are entire human societies, ranging from hunter-gatherer tribes to modern economies~\cite{wolpert2025,hayek2013use}.

\item \textbf{Computations by single cells in multi-cellular organisms.}
There are also many examples of \textit{single} biological organisms that are commonly viewed as performing computation,
even those lacking a nervous system. One canonical example is genetic regulatory
networks~\cite{Daniels18,Davidson2010},
which are often formulated as extensions of Boolean circuits, with recurrence allowed. Another example is ``chromatin computation'', which involves the feedback system of the expressed genes in a cell nucleus, creating proteins that then change which genes are expressed in that same cell nucleus~\cite{prohaska2010}. Even simple
ribosomes, translating RNA into amino acid sequences, have been viewed as a simple kind of computer~\cite{Kempes2017}.
See~\cite{Nicolau2016} for
other examples along these lines.

\item \textbf{ Canonical examples of non-constructed computers in philosophy of science}.
There has also been some semi-formal work in the philosophy of science on what criteria should be used to identify the computation done by a given dynamic system. For the most part,
this work does not distinguish between constructed and non-constructed computers. It also does not consider the broad range of kinds of (non-constructed) computers discussed above. Specific topics addressed in this literature range from (ontic) pancomputationalism to the role of counterfactuals to the physical Church-Turing thesis. 
See \cite{piccinini2021computation} for a review.
\end{itemize}

Next, we describe several important classes of dynamic systems that have been used in the literature to investigate
the relationship between dynamic systems, broadly construed, and computational machines. Since these 
dynamic systems are purely abstract,  strictly
speaking, they are constructed computers. Nonetheless, they are important to understand since they
illustrate many of the central issues in how dynamic systems can be seen as performing computation.
 
\begin{itemize}

\item \textbf{Systems far from thermodynamic equilibrium}. In the real world, essentially all computers that complete a calculation in finite time must operate far from thermodynamic equilibrium. In the last few decades,  a set of powerful theorems concerning the thermodynamics of such systems has been derived by applying recent results from stochastic thermodynamics to analyze idealized computational systems. 

These results from stochastic thermodynamics include various
thermodynamic speed limit theorems \cite{shiraishi2018speed,vu2023topological}, thermodynamic uncertainty relations \cite{barato2015thermodynamic,horowitz2020thermodynamic,van2022unified}, and first-passage time entropy production
results  \cite{PhysRevLett.119.170601,PhysRevLett.125.120604}. All of these results
provide formal bounds to the evolution of every
physical dynamic system, and so, in particular, these results apply to all 
physical computational systems, whether evolving stochastically or 
deterministically~\cite{Manzano2024prx,yadav_circuits_2025,yadav_rasp_2025,yadav2024minimal_channels}. 
The results in this body of work are sometimes called
``stochastic thermodynamics of computation''\cite{wolpert2019stochastic,wolpert2024computational}. Recent work 
in the stochastic thermodynamics of computation
has analyzed thermodynamic costs of TMs \cite{kolchinsky2020thermodynamic}, computation with circuits \cite{wolpert2020thermodynamics,yadav_circuits_2025}, and discrete finite automata \cite{Ouldridge_2023}, 
even when there is absolute irreversibility \cite{Manzano2024prx}. Many topics remain open \cite{wolpert2024computational}.

\item \textbf{Cellular automata}. 
Cellular automata (CA) are a class of 
dynamic systems that involve grids (typically infinite, and often one-dimensional) of ``cells'', each of which has its own small state space (often just a single bit)~\cite{von1966theory}. 
In each (discrete time) step of a CA, all of the cells update their state based on their earlier state and that of the cells in their immediate neighborhood on the grid. Typically, the rule for such updates is the same for all cells in the CA.
Therefore, this rule specifies a dynamic system whose state space is the joint state of all of its cells (sometimes called
a ``configuration"). 

CA can exhibit extremely rich types of behavior even for simple, homogeneous rules. As a result, they have been considered prototypical systems to investigate the possibility of the emergence of artificial life~\cite{LANGTON1986120,johnston2022conway}. 
Adopting a computer science paradigm instead, one would like to view any given
CA as a computational machine. The idea would be to identify the initial joint state of all the cells as the input to the computational machine,
while the subsequent dynamics of the joint state represent the computation itself \cite{GREIF2024126}. 

Crucially, though, there is no \textit{a priori} specification of how the configurations of a CA encode the 
inputs and outputs of a computation (never mind how such configurations at intermediate times
encode the intermediate states of a computational machine). In this, CAs are a canonical example of a dynamic system without any interpretation as a computer that we have chosen ahead of time.

That being said, there has been some quite fruitful work investigating the connection between CA and computation, where one chooses such an encoding ahead of time and then sets the initial configuration accordingly. This line of work stretches back at least to \cite{wolfram1984cellular}.
In that and subsequent work, Wolfram argued that a CA running ``Rule 110'' (using his terminology) is Turing complete \cite{WOLFRAM19841} when the
number of cells is countably infinite. This was later formally proven by Cook \cite{cook2004universality}. The deep relation between TMs, CA, and also both formal systems and Godel's incompleteness theorem was extensively discussed in  \cite{PROKOPENKO2019134}. This deep relation was more
recently used to define a ``biological arrow of time'' \cite{Prokopenko_2025}. The relation between CA and TMs has also been recently studied in the
specific case of fungi \cite{schumann2024}.

\item \textbf{Symbolic dynamics}. 
Deterministic dynamics of infinite strings of symbols is sometimes called ``symbolic dynamics''. It has been well studied in computer science, since many symbolic dynamics systems provide simplified descriptions of more complex computational processes. 

Indeed, note that the direct definition
of a symbolic dynamic system can be seen as a TM with the head removed, so that the dynamics of the
infinite tape/symbol sequence does not depend on the state of some such head. Despite this,
one can prove that the ``generalized shift map'' is computationally universal, i.e., as powerful
as arbitrary TMs 
\cite{moore1991generalized}. (See also~\cite{cardona2021constructing},
and \cite{gonzalez2025topological}, which is appearing in this special issue.) The proof of this result 
was soon followed  by closely related studies on topological dynamics \cite{KURKA1997203}
and computational universality of arbitrary symbolic dynamics \cite{delvenne2005computational}.

\item \textbf{Microstate dynamics of systems whose macrostate dynamics are constructed computers}.
There has been some very intriguing research considering the empirically observed low-level dynamics of real-world digital computational systems whose high-level dynamics implement digital programs. A natural research question concerning such systems is whether there is a (almost) bijective map between the ``kind of program'' being implemented and some broadly defined set of properties of the associated digital dynamic system. If there is, that raises the question of whether we can extend that bijection to infer what ``programs" are implemented by arbitrary dynamic systems, based on the properties of their dynamics. For some examples, see \cite{garland2011predicting} and  \cite{alexander2010}.
\end{itemize}

\section{Emulating computational machines with non-constructed computers}
\label{sec:formalization}

As described above, our goal in this paper is to investigate non-constructed computers, which
are dynamic systems where we must ``infer'' the decoding of some of the physical variables of that
system into the logical variables of a computational machine, so that we view the dynamic system
as implementing that machine. 
To address this issue in its full generality, we need to formalize precisely what we mean by such ``decoding''. In other words, we need a map $\varphi^{-1}$ taking (a coarse-graining of) an arbitrary dynamic system $A$ 
(either in discrete or continuous space and/or time) to a computational machine ${B}$
 such that the dynamics of  $A$ can be mapped to the dynamics of $B$.

\subsection{Formal definition of emulating a computational machine with a dynamic system}
\label{sec:formal_def}
First, we adopt the common definition that a \textbf{dynamic system} is any triple $(X, T, f)$ where:
\begin{enumerate}
\item $X$ is a set whose elements we call ``states'';
\item $T$ is a set of (real-valued) times;
\item $f: T \times T \times X \mapsto X$ 
\noindent
such that for all $x_t \in X$, for all $t,t',t'' \in T$ such that $t'' > t' > t$:
\begin{enumerate}
\item 
\label{item:iva}
$\qquad f(t, t'', x_t) = f(t', t'', f(t, t', x_t))$
\item  $\qquad f(t, t, x_t) = x_t$
\label{item:ivb}
\end{enumerate}
\end{enumerate}
The requirement of~\cref{item:iva} expresses the fact that we can interpret $f$ as the evolution operator of a system
whose dynamics is deterministic, and therefore Markovian. In contrast, the requirement of~\cref{item:ivb} simply expresses the fact
that the system cannot be in more than one state at once.
As shorthand, for all $A \subseteq X$, write $f(., ., A)$ for the map taking $A$ to the set $\{f(., ., a) : a \in A\}$.
We will say that a dynamic system  $(X, T, f)$ is \textbf{reversible} if for all $t, t' > t, x \in X$,
\begin{eqnarray}
\not\exists x' \ne x \, : \, f(t, t', x) = f(t, t', x')
\end{eqnarray}

Next, we need to define the computation that we wish $(X, T, f)$ to implement/emulate. Such a computation
is itself a dynamic system of course, just evolving over a space of ``computational variables'' $Y$ rather than over an arbitrary $X$. 
Moreover, in general, the dynamics of more than one $x \in X$ will emulate the evolution of a single element in $Y$.
(In other words, the emulation will be coarse-grained.) Accordingly,
we say that the dynamic system $(Y, T, g)$ is \textbf{emulated} by the dynamic system 
$(X, T, f)$ if there is a set-valued function $\varphi$ from elements in $Y$ to non-overlapping subsets of $X$ such that
\begin{enumerate}
\setcounter{enumi}{3}
\item 
\label{item:iv}
for all $t,t' \in T$, $y \in Y$,
\begin{eqnarray}
f(t,t',\varphi(y)) \subseteq \varphi(g(t,t',y))
\end{eqnarray}
\end{enumerate}	
\noindent	
We will abuse notation and write $\varphi^{-1}(x) := y$ for all $x \in \varphi(y)$. In the sequel, we will sometimes refer to the dynamical
system  $(Y, T, g)$ as a ``computational machine''.

We will sometimes refer to $\varphi^{-1}$
as the  \textbf{decoding} function, with $\varphi$ being the \textbf{encoding} function. 
Reflecting the roles usually played by the two dynamic systems, when the dynamic system 
$(Y, T, g)$ is time-reversible we will say that it is \textit{logically reversible}, and when  the dynamic system 
$(X, T, f)$ is time-reversible we will say that it is \textit{physically reversible}. (See \cite{piccinini2010computation} and references therein.)

Note that different $y$ get mapped by $\varphi$ to non-overlapping subsets of $X$, i.e., 
to different bins in a coarse-graining of $X$. Although in general it is not required, in the rest of this paper, we assume that the dynamics given by $f$ over these coarse-grained bins
is single-valued. To formalize this, write 
the coarse-grained bin containing any $x$ as $\mathscr{X}(x)$. So we
implicitly assume that for all $x, x' : \mathcal{X}(x) = \mathcal{X}(x')$, and for all $t, t' > t$, 
\begin{eqnarray}
\mathcal{X}[f(t, t', x)] = \mathcal{X}[f(t, t', x')]
\label{eq:3}
\end{eqnarray}
Combining this requirement with~\cref{item:iv}
means that we can follow a three-step procedure to determine how 
$g$ will evolve any $y \in Y$ from $t$ to $t'$. First, we pick any $x$ in the associated subset $\varphi(y) \subseteq X$. Next we
evolve that $x$ from $t$ to $t'$ using $f$. Then we find the (unique) $y'$ such that $\varphi(y')$ contains that evolved $x$.
We can write this three-step procedure as

\begin{eqnarray}
\varphi^{-1}f(t,t',\widehat{\varphi}(y)) = g(t,t',y)
\end{eqnarray}
where $\widehat{\varphi}(y)$ is a single-valued function from $Y$ to $X$ that is arbitrary,
so long as for all $y \in Y$, $\widehat{\varphi}(y) \in \varphi(y)$.

One notable peculiarity of emulation is how time arises in the two
dynamic systems. Suppose the dynamic system $(X, T, f)$ is a real physical
process while $(Y, T, g)$ is some abstract computational
machine, e.g., one of the machines in the Chomsky hierarchy. Our definition of emulation implicitly treats ``time'' in that computational machine as identical to ``time'' in the physical dynamic system.
At the cost of more complicated notation, we could extend the definition of 
emulation to allow the time that $f$ takes to finish its 
emulation to differ from the time that $g$ takes to perform the original map, 
i.e., so that the times $t$, $t'$ on the left-hand side of~\cref{item:iv} are not equal to the times $t$, $t'$ on the right-hand side, but rather are given by a function of those two times on the right-hand side. However, there is no
reason to consider such notation in this paper.

Note as well that it is straightforward to modify~\cref{item:iv} to allow nondeterministic computational machines
(as defined in computer science theory), by allowing $\varphi^{-1}$ to be a many-valued function. This would
mean that the sets $\varphi(y), \varphi(y')$ might overlap for certain pairs $y, y' \ne y$.
(We could alternatively capture nondeterministic computation by modifying the definition 
of $f$ to be many-valued.) Similarly, it is straightforward to modify~\cref{item:iv} to include 
stochasticity in $g$ (as in probabilistic TMs, stochastic finite automata, etc.).
As a final remark, note that the  definition of emulation given above is related to the concept of ``simulation'' in the theory of state transition systems,
albeit quite loosely.

\subsection{Formal definition of the central question of this special issue}

We can now formalize the topic of this paper as the following, \textit{central question}:

\begin{quote}
\textit{What is the set of all dynamic systems $(Y, T, g)$ that are emulated by any
given (arbitrary) dynamic system $(X, T, f)$?}
\end{quote}

Note that this central question is not the same as the question of whether a \textit{specific} 
dynamic system $(Y, T, g)$ is emulated by 
some other \textit{specific} dynamic system $(X, T, f)$.  Such variants of our central
questions have been investigated in the literature, mostly for the case where $(Y, T, g)$ is a universal Turing 
machine~\cite{wolfram1984cellular,moore1991generalized,moore1990unpredictability,lloyd1992any,lloyd2006programming,cardona2021constructing,gonzalez2025topological,shiraishi2021undecidability,cubitt2015undecidability,pour1982noncomputability}, or an extension of one~\cite{delvenne2005computational,davis1956note}.
This special case is discussed further below, in~\cref{sec:TM}.

In general, we (and the other contributors to this focus issue) are interested in these issues, subject to certain
constraints. For example, we might require that $T = \ZZZ$, $Y = \BBB^*$, and $(Y, T, g)$ is a machine in the Chomsky hierarchy.
A second example is where $f$ is an evolving physical system, and
we require $f$ to reach a physical state we interpret as a ``halting condition'' 
while using less than some bound on (physical) time. A related,
third example is where $f$ is again an evolving physical system, where
we might require $f$ to reach a physical state we interpret as a ``halting condition'' 
while using less than some bound on (physical) free energy. (Note that the second and third examples
differ only in what resource cost is of interest.)

Another kind of constraint that is very often imposed on our central question, even though it is almost never stated explicitly, 
is that no non-trivial computation is allowed to be ``hidden'' in the map $\widehat{\varphi}$.
Intuitively, very often we want to ensure that the dynamics of the computational machine being emulated is not hidden in the encoding function from the initial state of that machine to the initial state of the dynamic system
that will emulate it. A similar, more extreme constraint is that we do not want the initial state $\widehat{\varphi}(y)$ to already be the final result of the computation, $g(t, t', y)$. After all, if that were the case, it would mean that
$f(t, t', \widehat{\varphi}(y))$ only needs to implement the identity map. For similar reasons, we do not want any non-trivial computation to be ``hidden'' in the decoding function $\varphi^{-1}$.

A natural way to enforce such constraints
is to bound the time complexity or space complexity (in computer science's sense of the terms) of
any universal TM that implements the maps $\widehat{\varphi}$ and $\varphi^{-1}$~\cite{sipser2012introduction,arora2009computational}.
To illustrate this, consider the case where  $T$ is the set of all natural numbers. Suppose as well that
both $X$ and $Y$ are countably infinite (which, if $(X, T, f))$ is a real classical dynamic system, would mean that $X$ is a coarse-graining of phase space). So the state spaces of both dynamic systems can be represented as (a subset of)
$\mathbb{B}^*$, the set of all finite bit strings. 

Next, suppose we interpret
$(Y, T, g)$ as a universal TM, where $t \in T$ increases by $1$ for each iteration of 
of the TM, and similarly for $(X, T, f)$. This allows us to
characterize $f(t, t', x)$ as ``polynomial time'' if its value is given by the output of a
universal TM that runs for a time $t' - t$ that is a polynomial function
of the binary representation of the length of $x$ (or alternatively, of the
binary representation of $y = \varphi^{-1}(x)$). We can similarly
characterize  $f(t, t', x)$ as ``polynomial space'' if its value is given by the output of a
universal TM that uses a number of bits to complete which is a polynomial function
of (the binary representation of) the length of $x$ (or alternatively, of the
binary representation of $y = \varphi^{-1}(x)$).

Often in such a scenario, we have no  \textit{a priori} reason to require $\widehat{\varphi}$,
$f(t, t', x)$ etc., to be polynomial time and/or polynomial space. Even if we
accept the Church-Turing thesis and so forbid $f$ from being super-Turing, it might still
have super-polynomial computational power, e.g., being able to compute functions in \textbf{EXP} in
polynomial time.

However, as a practical matter, in real-world applications we typically want $\widehat{\varphi}$ to be 
easily computable, e.g., in polynomial time.
After all, if it's worse than polynomial time, then we would not be able to initialize the emulating dynamic
system $(X, T, f)$ in a timely manner if the initial bit string $y \in Y$ whose evolution we wish to emulate is too long.
Moreover, if we impose the constraint that  $\widehat{\varphi}$ 
be polynomial time and so polynomial space, it ensures that not much computation is hidden in $\widehat{\varphi}$. 

Similarly, again as a practical issue, we typically want $\varphi^{-1}$ to be easily 
computable (e.g., in polynomial time), in order to ``read out'' the state of the dynamic system $f$ at the time $t'$ as the result of the computation $g$. If we impose such a constraint, it ensures that not much
computation is hidden in $\varphi^{-1}$, see \cite{delvenne2009universal}. So by imposing these constraints on $\varphi^{-1}$ and $\widehat{\varphi}$, respectively, we ensure that neither is more computationally powerful than a polynomial 
computational machine $(Y, T, g)$.\footnote{As an aside, note that imposing such a constraint on $\widehat{\varphi}$ 
is closely analogous to the requirement
that a reduction (in the computer science sense of the term) from one computational machine to another
be polynomial time. In computer science, though, it is implicitly assumed that there is no
need to consider the time complexity of the analog of $\varphi^{-1}$, which
corresponds to inverting the reduction, to get back the answer to the original
decision problem. In fact, inverting functions is the focus of a different body of research in
computer science theory, involving one-way functions, cryptography, etc.~\cite{arora2009computational}.}

One downside of this approach to enforcing the constraints is that often, if $g$ is more computationally powerful than polynomial time,
we would want to allow $\widehat{\varphi}$ and/or $\varphi^{-1}$ to match $g$'s power, rather than
be forced to be weaker. One way to extend the approach to handle this scenario 
is to only require that $\widehat{\varphi}$ and/or $\varphi^{-1}$ be (polynomial time) reducible to $g$. 
For the case of $\widehat{\varphi}$, this would mean that there exists a polynomial-time TM $K$ mapping all ($t, t' > t, y)$ to a pair $(y^*, t^* > t)$ such that $\widehat{\varphi}(y) = g(t, t^*, y^*)$ and such that $K(t, t', y)$  completes in a time that is a polynomial function of the length of the binary representation of 
$(t, t', y)$~\cite{sipser2012introduction,arora2009computational}. (The case of $\varphi^{-1}$ would
be defined similarly.)

\section{Illustrations of the formal definition of dynamic systems emulating computational machines}
\label{sec:illustrations}




%


In this section, we briefly summarize some additional examples of dynamical systems that emulate computers (or seem to),
beyond those already presented in~\cref{sec:examples}. Our goal with these examples is to illustrate our definitions in~\cref{sec:formal_def}.

In all of these examples, we present an overview of how to establish that the dynamics of the appropriately instantiated physical systems is equivalent to a TM,
under the appropriate interpretation of its states in terms of logical variables (i.e., for the appropriate
decoding function). With some caveats, described below, these equivalences mean that the associated physical systems
are  ``Turing-complete'', i.e., capable of performing any computation that a TM can. 

In order to establish such Turing-completeness, the typical strategy is not to show directly the equivalence between the dynamical system and the TM. Usually, it is more convenient to construct a mapping to a system that is well-known to be Turing-complete (systems like the lambda-calculus, generalized recursive functions, register systems, Post canonical systems, or cellular automata, just to name a few). This is the strategy in the examples mentioned below.

\subsection*{Billiard ball model of computer}
One of the earliest attempts to build a computer that operated purely
mechanically was proposed by  Fredkin and Toffoli \cite{Fredkin1982}. They introduced the
concept of \emph{conservative logic}. Conservative logic is designed to respect both
physical and {logical reversibility}. It is also designed to respect
\emph{composition}, which can be related to assumption (iii) of the definition above. 
Specifically, they considered a model of balls with a fixed non-zero radius initially positioned and moving 
in a suitable fashion in a properly configured container (i.e., one containing appropriately angled walls).
They are mapping that system in such a way that the container configuration represents a specific 
logically reversible Boolean circuit, while the presence/absence of a ball in a specific position represents the state of a gate of that circuit. 
They then showed that, supposing collisions between the balls and between them and the walls of the container are purely elastic, this system could implement any Boolean circuit.
In this sense, their billiard ball model is computationally universal  \cite{sipser1996introduction}. 

We can identify the quantities in this model of a computer with the quantities in~\cref{sec:formal_def} as follows:
\begin{itemize}
    \item $X$  is an appropriate finite subset of the (classical) phase space of $N$ hard balls, $\Gamma$. We denote the coarse-grained space as $\underline{\Gamma}$.
    \item $f$ is the discrete-time dynamics through $\Gamma$ assuming they evolve according to frictionless kinetics with perfectly
elastic collisions. $f$ also encodes the container that lets the balls elastically reflect the container walls when they hit them. 
    \item $Y$ is the Cartesian product of an integer-valued discrete-time counter and $\BBB^N$, where $\BBB := \{0, 1\}$ and $N$ is the number of Boolean gates in the Boolean circuit that is being implemented by the billiard ball computer. 
    \item $g$ is specified by the dynamics of the (logically reversible) Boolean circuit, i.e., the (counter-dependent) 
discrete-time map setting the joint state of an associated subset of $N$ of the 
binary gates, based on the just-computed values of each gate's parents in the circuit.
    \item The domain of $\varphi^{-1}$ is the Cartesian product of  $\underline{\Gamma}$ and the value of the counter, $i$.  The function $\varphi^{-1}$ maps
any such element to the joint state of the subset of $N$ of the gates in the Boolean circuit that was just set after the $ i$'th iteration of the gates in the circuit. 
Note that the subset of the $N$ gates uniquely specifies the ending joint state of the output gates of the circuit. 

We assume that the dynamic system $(X, T, f)$ together with $\varphi$ (and therefore 
together with $\underline{\Gamma}$) obeys Eq.~\ref{eq:3}. 
At the initial value of the
counter, $Y$ includes the input gates, and at the final value of the counter, it includes the output gates.
\end{itemize}

 \subsection*{Computation with chemical reaction networks} 
 Another example of a  physical system that is capable of emulating a computer is a chemical reaction network (CRN), i.e., a finite set of coupled chemical reactions defined stoichiometrically, e.g., 
$$A+B \rightarrow C+D.$$ 
Appropriate CRNs can encode a broad set of computational
functions, where the input is the initial concentrations and the output is the final concentrations after the set
of reactions equilibrates.  It has been shown that non-inhibitory CRNs in which reaction rates can vary arbitrarily over time can encode continuous, piecewise linear functions \cite{chen2023rate}. On the other hand, inhibitor CRNs (iCRNs) that consist of reactions of the form 
$$A + B \stackrel{/ I}{\rightarrow}C+D,$$
where the reaction happens if reactants$A$ and $B$ are present and the inhibitor $I$ is absent,
can compute any computable function $f: \mathbb{N} \rightarrow \mathbb{N}$ \cite{doty2024}. This is shown by mapping the iCRNs to the register machines, which are known to be Turing-complete
\cite{minsky1967computation}.

We can identify the quantities in this model of a computer with the quantities in~\cref{sec:formal_def} as follows:
\begin{itemize}
\item $X$ is the set of concentrations of the chemical compounds used in the CRN, the initial concentrations correspond to the input variables, the final (i.e., the equilibrium) concentrations correspond to the output variables.
\item $f$ is the chemical reaction network, i.e., the list of reactions that are possible
\item $Y$ is the set of registers $R_i$ in the register model
\item $g$ is the set of instructions in the register model (e.g., the increment function \verb|INC| or decrement function \verb|DEC|)
\item $\varphi$ describes the mapping from the concentration operations through the iCRN to the set of instructions (i.e., the algorithm) in the register machine
\end{itemize}
    
\subsection*{Fluid computers}
We end with a discussion of some types of dynamical systems that do \textit{not} emulate TMs, in the formal sense of our definitions, even though they do have a certain kind of ``Turing-completeness''.

Ever since \cite{moore1991generalized}, several researchers have 
raised the question of whether fluid flows are capable of performing (universal) computation. Tao 
suggested \cite{tao2019} there might be a connection between potential 
blow-ups of Navier-Stokes equations, Turing completeness, 
and such fluid computation. In a recent paper \cite{cardona2021constructing}, the authors 
constructed a TM from a dynamical system described by the Euler flow on Riemann $\mathbb{S}^3$ by identifying the flows with generalized shift maps that are equivalent (``conjugate'') to a TM. The way they did this can be summarized as follows. First, they construct a compact $3$-dimensional manifold where Euler flows are defined, i.e., where the set of stationary vector fields $\mathbf{u}(x)$ that fulfill the following conditions
$$\nabla \mathbf{u} = 0 \quad\mathrm{(incompressibility)}$$
$$(\mathbf{u} \cdot \nabla) \mathbf{u} = - \nabla p \quad \mathrm{(Euler \ equation)}$$
is well-defined.
The streamlines (also called \emph{trajectories}) of $\mathbf{u}$ $x_\mathbf{u}$ ($\dot{x}_{\mathbf{u}} = \mathbf{u}(x_\mathbf{u})$) 
are identified with time evolution on the manifold. The conjugacy with TMs 
is established through showing the equivalence between the Euler flow and the Generalized shift map, which was shown to be conjugate to a TM in earlier work \cite{moore1991generalized}. 

A small variant of this group of authors also showed the existence of ``Turing-complete'' flows in other compact manifolds \cite{CARDONA2023109142} as well as in Euclidean spaces \cite{CARDONA202350}.
Furthermore, in a series of recent studies, the authors show that the flow on a smooth bordism of a vector field with good local properties can simulate any Turing-computable function  \cite{gonzalez2025topological}.  Finally, they construct Turing-complete steady flows of Navier-Stokes fields \cite{dyhr2025}. A recent discussion can be found in Ref.~\cite{González-Prieto_2025}.

In a recent study~\cite{cardona2025towards}, this and related approaches to emulating computation
in dynamical systems were
summarized with the following definition of (what the authors call) ``Turing-complete'':

\begin{quote}
``A dynamic system $X$ on the topological space $M$ is Turing complete if there exists a universal TM $T_U$
such that for each initial configuration $c$ of $T_U$, there exists a (computable) point $p_c \in M$ and a (computable) open
set $U_c \subset M$ such that $T_U$ halts with input $c$ if and only if the positive trajectory of $X$ through $p_c$ intersects $U_c$.''
\end{quote} 

\noindent Although dynamical systems that have this ``fluid computer Turing-completeness'' (FCTC) are  
similar in some aspects to  the dynamical systems that emulate UTMs (in the formal sense
defined in~\cref{sec:formalization}), there are some important differences.

%
%

Often, we are not interested in whether a given TM halts for a given input. Rather,
we want to know the \textit{output} of a TM when it does halt, for such a given input.
To use an FCTC to decide even just whether a TM $T$ that has input $I$ halts with \textit{some specific} output
$O$, one needs to encode $I$ and $O$ both in 
  $p_c$.
Given some such encoding, one could use a countably infinite set of FCTCs, each run on a different $p_c$, to perform 
the desired computation of the actual output $O$. For example, one could do this using
the ``dovetailing'' procedure of CS theory. However, this procedure would require a second dynamical system to sift among that infinite set of FCTCs, watching which has halted when, and what output $O$ is encoded
in the $p_c$ of that FCTC if and when it does halt.~\footnote{In a dovetailing procedure, one would run the dynamical system simultaneously, in a staggered manner, for more and more output strings $O$, so that if the TM $T$ halts with input $I$, so would the dovetailed instances
of the dynamic system, and furthermore, the instance of the dynamic system that halts would be the one specifying the precise output $O$ produced by $T$ on input $I$. (One might think it would suffice to just run a single FCTC iteratively, for every
possible output $O$. However, this would have the problem that for some given $T, I$, where $T(I) = O$,
the dynamical system never halts
for some output $O'$ that it's run on (in the sense that it goes through the point $U_{I, O, T}$) before it's ever run on the correct output, $O$.
The dovetailing procedure avoids this problem by running iteratively more and more instances of the dynamical system at the same time, 
just for different inputs, in a staggered fashion~\cite{li2008introduction}. This dovetailing approach will result in one of the 
instances of the dynamical system halting with the correct output $O$, and no instances will halt with an
incorrect output.) })
No such dynamical system is considered in~\cite{cardona2021constructing}.

Note as well that even if the definition of an FCTC were extended somehow to have it
compute the output string when a given TM with a given input halts, FCTCs would still only implement the partial function of that TM. They provide no information about the full dynamics of that TM through its 
state space of instantaneous descriptions (IDs), as it evolves from one iteration of its update function to the next. 
In other words, they do not construct an analog of the map $\varphi$ central to our approach, taking an arbitrary state of the TM to a subset of $X$.
Nor does the definition of an FCTC involve a set $T$ of times, like those considered in our
approach. These differences between FCTCs and the kinds of systems we consider above
reflect the fact that the goals in the investigations of FCTCs are different from ours.

\section{Uncomputability in dynamic systems}
\label{sec:TM}
Many dynamic systems $(X, T, f)$
have been explicitly shown to have properties that are uncomputable, in that no TM that is guaranteed to halt for all inputs can compute those properties ahead of time \cite{pour1982noncomputability}. Recent work has discussed examples of such systems. Bausch et al. \cite{Bausch2021} proved that for a specific Hamiltonian
of a many-body quantum system, the task of determining the phase diagram is generally uncomputable. This result is based on the fact that the problem of deciding whether a many-body quantum system has a spectral gap is at least as hard as solving the Halting problem, and therefore undecidable
(i.e., the bit of whether there is a spectral gap is uncomputable) \cite{Cubitt2015}. 
Similarly, Shiraishi and Matsumoto \cite{shiraishi2021undecidability} showed that for a general quantum many-body system, the question of whether the system thermalizes is undecidable. Undecidability has not only been established for
quantum systems. For instance, Gonda et al. \cite{compositionality:14134} discussed whether a general spin model is equivalent to a TM. Additionally, Purcell et al. \cite{purcell2024chaitin} created a family of Hamiltonians that has a guaranteed single phase transition. Yet, the location of this phase transition is uncomputable since it is determined by the value of Chaitin's constant, which is known to be uncomputable. 

These kinds of results have motivated several authors to suggest that it might
be useful to study physical dynamic systems with more computational power than TMs \cite{SIEGELMANN1996461,pour2017computability}. These systems are typically called \textit{super}-Turing capable systems in the literature.

The concept of dynamic systems with super-Turing capability leads us to carefully consider the original 
form of the Church-Turing thesis. As recently discussed in \cite{copeland2018church}, there are at least three forms of the Church-Turing thesis. The strongest version called \emph{Super-bold Physical Church-Turing thesis} or \emph{Total Physical Computability Thesis}, assumes that every physical aspect of any real-world system must be Turing computable. While some of the aforementioned examples might point to the fact that this thesis might be too strong, it is open to debate whether the uncomputability problem is not just a mathematical artifact of the underlying theories (e.g., quantum mechanics) and whether it is rather not a constraint that the underlying physical theories should follow. 

On the other hand, there are at least two weaker versions of the Church-Turing thesis that are not so restrictive. The first one, the \emph{Church-Turing-Deutsch-Wolfram} thesis, proposes that every physical system can be simulated on a TM. An even more modest version, 
called the \emph{Physical computation thesis}, proposes that every function computed by any physical computing system is Turing-computable. 

To date, it is not known which (or even whether) any of the forms of the Church-Turing thesis is correct.
However, recent discussion suggests that the Church-Turing thesis should not be understood so much as being 
about the computability or super-computability of a physical system, but rather about whether we could ever confirm that a dynamic system we encountered has such capabilities. 

\section{Value of computation}

For many dynamic systems, both constructed and non-constructed, it might not be particularly important 
to identify the precise computation done by the dynamic system, 
nor the amount of computation that it does. Rather, it is the \textbf{value of computation} done
by such systems that is their most important property. Such a value of computation might be the most important quantity 
concerning the future evolution of the system. A typical example is (biological) {evolutionary systems}, in which the value 
of computation can be defined by the future dynamics of one or more fitness functions. 

To illustrate this in more detail, consider an agent that has some initial information about its environment, and which uses that information to \textit{compute} the best action to take on its environment, in order to increase the value of its {``viability''}.
For example, this viability may be the ability of the agent to
extract free energy from its environment, i.e., to adaptively respond to environmental perturbations that it senses, thereby
maximizing long-term reproductive fitness. The better the computation the agent does, the higher its viability will be after this subsequent interaction with its environment. Some preliminary work has been done in Refs.~\cite{kolchinsky2021,hartle2024work,hartle2025distributed}.

To make this concrete, suppose we have 
three time-steps, $t = \{t_1, t_2, t_3\}$, and 
an agent who comprises two variables, a memory variable,
$X^m$, and an action variable, $X^a$. We represent the agent's variables as $X=\{X^m,X^a\}$. For simplicity, we represent the state of the environment as a single variable, $Y$. 
The joint system's dynamics proceed as follows:

\begin{enumerate}
\item At time $t=t_1$, there is some statistical coupling between $X^m$ and $Y$, e.g., reflecting earlier observations by the
agent of variables in the environment.
    \item During the time interval $t \in (t_1, t_2]$, $X$ and $Y$ evolve independently of one another, i.e.,
$P(y_{t}, x_{t} | y_{t_1}, x_{t_1}) = P(y_{t} | y_{t_1})P(x_{t} | x_{t_1})$. We can also assume that $X^a$ does not change during that evolution. 
We identify the evolution of the agent's joint memory state during this time as its performing a ``computation'',
based on the initial value of its memory at the start of the interval. 
\item At the end of this interval,  at $t=t_2$, there is a conditional
distribution $P(X^a | X^m)$ that uses the state of
the agent's memory to set the value of their action variable.
\item Then, during $t \in (t_2, t_3]$, the agent's variable $X^a$ 
and environment are coupled via some (exogenously specified) interaction Hamiltonian $H_{int}(x,y)$ that was set to $0$ before
the interval
\item At $t=t_3$, that interaction Hamiltonian is set to $0$, i.e., it is removed. For completeness, we can
assume that $X^m$ does not change during that time interval $t \in (t_2, t_3]$.
\item At this point, the entire cycle can repeat, perhaps after an intervening interval in which the statistical coupling between
$X^m$ and $Y$ is reset.
\end{enumerate}

As in the semantic value of information framework \cite{kolchinsky_semantic_2018}, we are interested in the value of a viability function concerning the distribution $P(x_{t_3})$.
Examples of that function include the KL divergence from $P(x_{t_3})$ to the stationary state of $X$ under a rate matrix
governing its dynamics, or the KL divergence to the Boltzmann distribution of the
agent~\cite{WolpertKolchinsky2016,kolchinsky_semantic_2018}.  (Note the similarity of this setup to closed-loop control, in control theory,
just with unusual objective functions.)

In other situations, other kinds of viability are more directly of interest.
In particular, often the relevant type of viability involves \textit{embodiment constraints}.
As an example, often we are interested in the minimal amount of 
thermodynamic work required by the agent to go from the distribution over its possible states $(X_m, X_a)$
at $t_3$ to some desired target distribution over those states, before the cycle repeats. In this
case, the viability function would be (negative of) the amount of work expended to effect that
change in the distribution. Crucially though, the agent must implement that change in distributions while subject to some set of constraints on rate matrices and/or Hamiltonians governing the dynamics of $X$, constraints
which reflect limitations of the agent. Those are the embodiment constraints.



As a variant, in some situations it makes sense to consider viability functions that do not involve the amount of work required to go from one 
ending state distribution to another. For example, if the agents under consideration are biological 
organisms, the viability function could be the difference in the beginning and ending distributions of some particular non-energetic resource 
(like carbon, or phosphorus, or some such) rather than the amount of work. Such viability functions can be of interest
either where there are embodiment constraints or not, depending on the precise scenario that is being modeled.


In all of these scenarios, we would identify the dynamics from $X^m_{t_1}$ to $X^m_{t_2}$ as emulating some specific computation, i.e., that dynamics is the computation performed by the agent.
The precise form of the dynamics over $X^m$ would specify the precise computational
machine being emulated by the agent. The value of that computation would quantify
the ``power'' or ``complexity'' of that computation.


\section{Future Work}
\label{sec:future}


There are many avenues to be explored for future research. Some were
discussed above, in the main text. Here we list a few
others of particular interest.


\begin{itemize}
\item 
\label{item:i}
\textbf{Computers that interact with input and output systems} 

All constructed computers
crucially rely on there being at least two additional dynamic systems that they interact with:
\begin{enumerate}[label=(\Roman*)]
    \item[(I)] An input system to determine the initial state of the computer and/or an input (or stream of inputs) into that computer; 
    \item[(O)] An output system to observe the outputs of the computer.
\end{enumerate}
(The analysis of computers that receive streams of inputs was pioneered by Karp; see~\cite{karp1992line}.)

Unless it is accompanied by these extra systems, no human would be able to actually \textit{use} a particular constructed computer. However, just because it is accompanied by such input and output systems does not ensure that a given constructed computer is usable. 
For example, an input system that requires fixing an infinite number of digits in an arbitrary real number would be ruled out,
as (tautologically?) beyond human capability to construct. 

How does the problem of identifying computation in real-world dynamic systems
change if we expand it to involve triples of dynamic systems, including both (usable) input and output systems, in addition to a computer system that they are coupled with? How does the problem change if we combine (I) and (O) into a single system, an external environment that interacts with the computer’s outputs and determines its inputs?

\item \textbf{Dynamical systems as optimizers of an objective function.} 

One approach to inferring what computation is done by an arbitrary non-constructed dynamic system relies on knowing (or assuming) an objective function that the dynamic system optimizes. In particular, some research in neurobiology has imposed specific tasks on a biological system and then tried to use the system's (neurological) response to those tasks to infer the dynamic system's computational algorithm for achieving that task. This approach implicitly presupposes that the dynamics of the biological system are optimized to perform well at the specific task that the researcher is able to impose on the system. For example, if the biological system is an organism with a central nervous system and a large brain, this approach has relied on 
presupposing that the brain can perform optimally at the task of recording a memory~\cite{Kosse2019}, or at the task
of decomposing the form of a sensory stimulus after the brain has processed it into a form that matches the structure of the stimulus~\cite{bornschein2013v1}.

However, in the vast majority of non-constructed dynamic systems, we have no such objective function we can rely on to identify the computation performed by the system. Even in the context of neurobiology, often the fundamental goal an organism is addressing can remain elusive, never mind what computation they might be using to achieve such a goal. 

If we do restrict ourselves to systems where there is a clear objective function, what are the best methods for inferring from the change in the dynamics of the system when the task is imposed, exactly what computational algorithm is used to achieve the task?

\item \textbf{Quantifying the amount of computation emulated by a dynamical system rather than the precise computation} 

A perhaps simpler issue than identifying the precise computation emulated by a dynamic system $(X, T, f)$
would be to identify the ``amount of computation" done by that system. In general, this would mean we don’t need to consider an explicitly emulated dynamic system $(Y, T,g)$, or a map taking the dynamics given by $g$ to the dynamics given by $f$~\cite{li2025measuring}. 

As an illustration of this approach, we should mention the rich theory that quantifies the ``randomness'' of an infinite string
using the machinery of algorithmic randomness (i.e., prefix-free Kolmogorov complexity). The idea would be to suppose
that any given observed finite string $\omega$ of length $n$ is just the first $n$ symbols of an
infinite string, $\hat{\omega}$. The supposition is that the infinite string $\hat{\omega}$ is being
generated by some computational system, with noise added (e.g., due to observational limitations). 
The idea is to examine the substrings in $\omega$ to
estimate both the rate of ``how much computation'' was being done by the computational system,
and the rate of noise being added to generate $\hat{\omega}$.

For example, in the limit that $n \rightarrow \infty$, \textit{if the string were random in the Martin-L\"of sense}, then its randomness deficiency would approach a constant, i.e., the difference between the (prefix)
Kolmogorov complexity of the first $n$ symbols in the strings and the length of the string would
approach a constant as $n \rightarrow \infty$~\cite{li2008introduction,mota2013sophistication,downey2019computability}.
To generate any such maximally random string, a (prefix) universal TM
would require the maximal possible (rate of )``amount of computation''.

To instead measure the (rate of) computation performed by a dynamic system that does \textit{not}
saturate the upper bound, it is natural to hypothesize that the Kolmogorov complexity $K$ of the string $\omega$ of length $n$ obeys
\begin{eqnarray}
K(\omega) = \alpha n + c + O(1)
\end{eqnarray}
for some constants $\alpha, c$. (The constant $\alpha$ is sometimes called the ``Kolmogorov rate'' in the literature.) We could then suppose that $\alpha$ quantifies the fraction of the amount of computation in generating $\omega$,
i.e., the amount not due to noise. 

To use this idea in practice, we would want to estimate $\alpha$ from our data, i.e., by examining substrings
of our single finite string $\omega$. In turn, to make such an estimate, we would need to use a bound on Kolmogorov complexity,
since that quantity is uncomputable.
Standard choices for an (upper) bound are Lempel-Ziv codelength~\cite{ziv1978compression}, along with its many
variants. (Loosely related approaches have been used to estimate the complexity of biological systems \cite{ZENIL201632}, cellular automata \cite{estevez2015lempel}
or neuronal spike trains \cite{haslinger2010computational}.)

There are many variations of this approach that might be worth pursuing. Some would involve considering the rate of the conditional Kolmogorov complexity of the most recent symbol in the string, conditioned on the preceding substring as the string length grows. One might even consider quantities like the entropy rate~\cite{cover2006elements} rather than the Kolmogorov rate, especially in light of their tight relationship in the limit of large strings. Of course, yet another related set of approaches could be applied if our data consisted of multiple finite strings, rather than a single
one, so that we would focus on a distribution over strings and the associated expectation of the
amount of computation. 


\item \textbf{Higher-order Markov chain process and amount of computation} 

As another example of how to quantify the amount of computation done by a dynamic system, 
note that any real-world computer has lots of memory, reflecting (aspects of its) previous state. This means that the 
variables in the computer's core evolve in a way that is not first-order Markov. To illustrate this, suppose we look at only the flash RAM of a constructed computer, without considering what’s also stored on the disk, the dynamics of that flash RAM is a higher-order Markov process, or equivalently, a hidden Markov model (HMM).

This suggests that we measure ``how much computation" a given dynamic system $Z_t$ is doing by looking at the order of the Markov chain driving that dynamics, and/or the number of hidden states in a first-order HMM that is driving that dynamics $Z_t$. There are many ways we could do this. 
For example, we could consider the maximal delay $\tau$ such that the mutual information $I(Z_t ; Z_{t-1},\dots, Z_{t-\tau}) > \epsilon$ for some prefixed threshold $\epsilon$. Alternatively we could take the maximal $\tau$ such that $I(Z_t ;  Z_{t-\tau}) > \epsilon$. It is not a priori clear which approach is more beneficial, see~\cite{li2025} for some preliminary work along these lines. Alternatively, one can consider a delay-embedding of the process $Z_t$ and 
estimate the minimal embedding dimension necessary for the process to be linear. (Cf., Takens' theorem \cite{takens2006detecting}.)
	

\item 
\label{item:iii} \textbf{Hierarchical structure of non-constructed computers}

Many dynamic systems found in nature that we view as performing computation are physically distributed (often with a hierarchical, modular structure). Can we gain insight into the dynamic behavior of spatially distributed, heterogeneous dynamic systems in general by reformulating them in a first-principles way as performing computation?

\item \textbf{Dynamical systems as models of continual computers} 

Naturally occurring dynamic systems that we view as performing computation are very different from
 standard one-shot computers like those performed by the machines in the Chomsky hierarchy, by 
machines like CAs, by systems implementing
symbolic dynamics, etc. Such one-shot computers receive a single set of one or more inputs at a given time and then compute the associated output(s). 
In contrast, naturally occurring systems like the brain are \emph{continual computational systems} --- they are getting new inputs continually, at random
times, each input requiring a different computational task, and those tasks overlapping with one another in time. Physically, these continual
computational systems are not just open to the outside world once at initialization. Rather, they are always open, always being perturbed. They are
``embodied" in the 
real world~\cite{karp1992line,roscoe2010understanding,reisig2016petri}. 
Does it help us to understand how arbitrary real-world dynamic systems should be viewed as computers if we restrict attention to such open, embodied systems? 

%
%

\item \textbf{Collective computation}

Suppose one has a model of the dynamics of a system. To quantify ``collective computation, focusing on the collective component specifically, one can see how much the dynamics change if we sever all communication among the subsystems. This quantifies the
\textit{collective} aspect of the computation, per se. It doesn’t say what it is that’s being computed, of course.
\end{itemize}

These are just a few of the many open problems regarding the theory of non-constructed computers. Many of these problems are connected with the fact that computation is closely related and absolutely crucial for all living systems. Without computation, no living systems could exist. However, as we discussed above, many other dynamical systems are capable of computation. We believe that solving some of the aforementioned issues will be beneficial for understanding both living and non-living non-constructed computers.

\section*{Acknowledgements}
DHW would like to thank the Santa Fe Institute for support. JK acknowledges support from the Austrian Science Fund (FWF) project No. P34994 (grant-DOI: 10.55776/P34994). The research was also co-financed by the Austrian Research Promotion Agency (FFG) as part of the ESSENCSE project (873927) and the Austrian Federal Ministry for Climate Action, Environment, Mobility, Innovation \& Technology (BMK) as part of the CSH Postdoc Programme (GZ 2023-0.841.266)

\section*{References}

\bibliographystyle{unsrt}
\bibliography{bibliography.bib}

\begin{thebibliography}{100}

\bibitem{navlakha2011algorithms}
Saket Navlakha and Ziv Bar-Joseph.
\newblock Algorithms in nature: the convergence of systems biology and computational thinking.
\newblock {\em Molecular systems biology}, 7(1):546, 2011.

\bibitem{haken2012computational}
Hermann Haken.
\newblock {\em Computational Systems---Natural and Artificial: Proceedings of the International Symposium on Synergetics at Schlo{\ss} Elmau, Bavaria, May 4--9, 1987}, volume~38.
\newblock Springer Science \& Business Media, 2012.

\bibitem{Chu2018computation}
Dominique Chu, Mikhail Prokopenko, and J.~Christian~J. Ray.
\newblock Computation by natural systems.
\newblock {\em Interface Focus}, 8(6):20180058, 2018.

\bibitem{wolpert2024computational}
David~H Wolpert, Jan Korbel, Christopher~W Lynn, Farita Tasnim, Joshua~A Grochow, G{\"u}lce Karde{\c{s}}, James~B Aimone, Vijay Balasubramanian, Eric De~Giuli, David Doty, et~al.
\newblock Is stochastic thermodynamics the key to understanding the energy costs of computation?
\newblock {\em Proceedings of the National Academy of Sciences}, 121(45):e2321112121, 2024.

\bibitem{sipser1996introduction}
Michael Sipser.
\newblock Introduction to the theory of computation.
\newblock {\em ACM Sigact News}, 27(1):27--29, 1996.

\bibitem{arora2009computational}
Sanjeev Arora and Boaz Barak.
\newblock {\em Computational complexity: a modern approach}.
\newblock Cambridge University Press, 2009.

\bibitem{li2008introduction}
Ming Li, Paul Vit{\'a}nyi, et~al.
\newblock {\em An introduction to Kolmogorov complexity and its applications}, volume~3.
\newblock Springer, 2008.

\bibitem{meijers2021behavior}
Matthijs Meijers, Sosuke Ito, and Pieter~Rein Ten~Wolde.
\newblock Behavior of information flow near criticality.
\newblock {\em Physical Review E}, 103(1):L010102, 2021.

\bibitem{piccinini2021computation}
Gualtiero Piccinini and Corey Maley.
\newblock Computation in physical systems.
\newblock {\em Stanford Encyclopedia of Philosophy}, 2021.

\bibitem{cotler2024computationaldynamicalsystems}
Jordan Cotler and Semon Rezchikov.
\newblock Computational dynamical systems, 2024.

\bibitem{urai2022large}
Anne~E Urai, Brent Doiron, Andrew~M Leifer, and Anne~K Churchland.
\newblock Large-scale neural recordings call for new insights to link brain and behavior.
\newblock {\em Nature neuroscience}, 25(1):11--19, 2022.

\bibitem{furber2016large}
Steve Furber.
\newblock Large-scale neuromorphic computing systems.
\newblock {\em Journal of neural engineering}, 13(5):051001, 2016.

\bibitem{schuman2022opportunities}
Catherine~D Schuman, Shruti~R Kulkarni, Maryam Parsa, J~Parker Mitchell, Prasanna Date, and Bill Kay.
\newblock Opportunities for neuromorphic computing algorithms and applications.
\newblock {\em Nature Computational Science}, 2(1):10--19, 2022.

\bibitem{markovic2020physics}
Danijela Markovi{\'c}, Alice Mizrahi, Damien Querlioz, and Julie Grollier.
\newblock Physics for neuromorphic computing.
\newblock {\em Nature Reviews Physics}, 2(9):499--510, 2020.

\bibitem{deutsch1992rapid}
David Deutsch and Richard Jozsa.
\newblock Rapid solution of problems by quantum computation.
\newblock In {\em Proceedings of the Royal Society of London A: Mathematical, Physical and Engineering Sciences}, volume 439, pages 553--558. The Royal Society, 1992.

\bibitem{feynman1986quantum}
Richard~P Feynman.
\newblock Quantum mechanical computers.
\newblock {\em Foundations of physics}, 16(6):507--531, 1986.

\bibitem{nielsen2006quantum}
Michael~A Nielsen, Mark~R Dowling, Mile Gu, and Andrew~C Doherty.
\newblock Quantum computation as geometry.
\newblock {\em Science}, 311(5764):1133--1135, 2006.

\bibitem{siegelmann1995computation}
Hava~T Siegelmann.
\newblock Computation beyond the turing limit.
\newblock {\em Science}, 268(5210):545--548, 1995.

\bibitem{siegelmann1998analog}
Hava~T Siegelmann and Shmuel Fishman.
\newblock Analog computation with dynamical systems.
\newblock {\em Physica D: Nonlinear Phenomena}, 120(1-2):214--235, 1998.

\bibitem{maley2022and}
Corey~J Maley.
\newblock How (and why) to think that the brain is literally a computer.
\newblock {\em Frontiers in Computer Science}, 4:970396, 2022.

\bibitem{Chialvo2010}
Dante~R. Chialvo.
\newblock Emergent complex neural dynamics.
\newblock {\em Nature Physics}, 6(10):744--750, Oct 2010.

\bibitem{hesse2014self}
Janina Hesse and Thilo Gross.
\newblock Self-organized criticality as a fundamental property of neural systems.
\newblock {\em Frontiers in systems neuroscience}, 8:166, 2014.

\bibitem{Shew15595}
Woodrow~L. Shew, Hongdian Yang, Thomas Petermann, Rajarshi Roy, and Dietmar Plenz.
\newblock Neuronal avalanches imply maximum dynamic range in cortical networks at criticality.
\newblock {\em Journal of Neuroscience}, 29(49):15595--15600, 2009.

\bibitem{Brochini2016}
Ludmila Brochini, Ariadne de~Andrade~Costa, Miguel Abadi, Ant{\^o}nio~C. Roque, Jorge Stolfi, and Osame Kinouchi.
\newblock Phase transitions and self-organized criticality in networks of stochastic spiking neurons.
\newblock {\em Scientific Reports}, 6(1):35831, Nov 2016.

\bibitem{PhysRevLett.126.098101}
Leandro~J. Fosque, Rashid~V. Williams-Garc\'{\i}a, John~M. Beggs, and Gerardo Ortiz.
\newblock Evidence for quasicritical brain dynamics.
\newblock {\em Phys. Rev. Lett.}, 126:098101, Mar 2021.

\bibitem{clark2013whatever}
Andy Clark.
\newblock Whatever next? predictive brains, situated agents, and the future of cognitive science.
\newblock {\em Behavioral and brain sciences}, 36(3):181--204, 2013.

\bibitem{doi:10.1073/pnas.1912340117}
Richard~M. Shiffrin, Danielle~S. Bassett, Nikolaus Kriegeskorte, and Joshua~B. Tenenbaum.
\newblock The brain produces mind by modeling.
\newblock {\em Proceedings of the National Academy of Sciences}, 117(47):29299--29301, 2020.

\bibitem{shain2020fmri}
Cory Shain, Idan~Asher Blank, Marten van Schijndel, William Schuler, and Evelina Fedorenko.
\newblock fmri reveals language-specific predictive coding during naturalistic sentence comprehension.
\newblock {\em Neuropsychologia}, 138:107307, 2020.

\bibitem{chalk2018toward}
Matthew Chalk, Olivier Marre, and Ga{\v{s}}per Tka{\v{c}}ik.
\newblock Toward a unified theory of efficient, predictive, and sparse coding.
\newblock {\em Proceedings of the National Academy of Sciences}, 115(1):186--191, 2018.

\bibitem{gebicke2023computational}
Peter~J Gebicke-Haerter.
\newblock The computational power of the human brain.
\newblock {\em Frontiers in Cellular Neuroscience}, 17:1220030, 2023.

\bibitem{Khuong2016}
Ana{\"\i}s Khuong, Jacques Gautrais, Andrea Perna, Chaker Sba{\"\i}, Maud Combe, Pascale Kuntz, Christian Jost, and Guy Theraulaz.
\newblock Stigmergic construction and topochemical information shape ant nest architecture.
\newblock {\em Proceedings of the National Academy of Sciences}, 113(5):1303--1308, 2016.

\bibitem{HEYLIGHEN20164}
Francis Heylighen.
\newblock Stigmergy as a universal coordination mechanism i: Definition and components.
\newblock {\em Cognitive Systems Research}, 38:4--13, 2016.
\newblock Special Issue of Cognitive Systems Research -- Human-Human Stigmergy.

\bibitem{doi:10.1126/science.ade1702}
Shihao Dong, Tao Lin, James~C. Nieh, and Ken Tan.
\newblock Social signal learning of the waggle dance in honey bees.
\newblock {\em Science}, 379(6636):1015--1018, 2023.

\bibitem{Salman2024}
Muhammad Salman, David Garz{\'o}n~Ramos, and Mauro Birattari.
\newblock Automatic design of stigmergy-based behaviours for robot swarms.
\newblock {\em Communications Engineering}, 3(1):30, Feb 2024.

\bibitem{Nichol_2024}
Peter~B. Nichol.
\newblock Why is stigmergy a good platform for swarm intelligence?, 2024.

\bibitem{theraulaz1999brief}
Guy Theraulaz and Eric Bonabeau.
\newblock A brief history of stigmergy.
\newblock {\em Artificial life}, 5(2):97--116, 1999.

\bibitem{BALLERINI2008201}
Michele Ballerini, Nicola Cabibbo, Raphael Candelier, Andrea Cavagna, Evaristo Cisbani, Irene Giardina, Alberto Orlandi, Giorgio Parisi, Andrea Procaccini, Massimiliano Viale, and Vladimir Zdravkovic.
\newblock Empirical investigation of starling flocks: a benchmark study in collective animal behaviour.
\newblock {\em Animal Behaviour}, 76(1):201--215, 2008.

\bibitem{Adamatzky03042015}
Andrew~Adamatzky and.
\newblock Slime mould computing.
\newblock {\em International Journal of General Systems}, 44(3):277--278, 2015.

\bibitem{wolpert2025}
David~H. Wolpert and Kyle Harper.
\newblock The computational power of a human society: a new model of social evolution, 2025.

\bibitem{hayek2013use}
Friedrich~August Hayek.
\newblock The use of knowledge in society.
\newblock In {\em Modern understandings of liberty and property}, pages 27--38. Routledge, 2013.

\bibitem{Daniels18}
Bryan~C. Daniels, Hyunju Kim, Douglas Moore, Siyu Zhou, Harrison~B. Smith, Bradley Karas, Stuart~A. Kauffman, and Sara~I. Walker.
\newblock Criticality distinguishes the ensemble of biological regulatory networks.
\newblock {\em Phys. Rev. Lett.}, 121:138102, Sep 2018.

\bibitem{Davidson2010}
Eric~H. Davidson.
\newblock Emerging properties of animal gene regulatory networks.
\newblock {\em Nature}, 468(7326):911--920, Dec 2010.

\bibitem{prohaska2010}
Sonja~J. Prohaska, Peter~F. Stadler, and David~C. Krakauer.
\newblock Innovation in gene regulation: The case of chromatin computation.
\newblock {\em Journal of Theoretical Biology}, 265(1):27--44, 2010.

\bibitem{Kempes2017}
Christopher~P. Kempes, David Wolpert, Zachary Cohen, and Juan P{\'e}rez-Mercader.
\newblock The thermodynamic efficiency of computations made in cells across the range of life.
\newblock {\em Philosophical Transactions of the Royal Society A: Mathematical, Physical and Engineering Sciences}, 375(2109):20160343, 2017.

\bibitem{Nicolau2016}
Dan~V. Nicolau, Mercy Lard, Till Korten, Falco C. M. J.~M. van Delft, Malin Persson, Elina Bengtsson, Alf M{\aa}nsson, Stefan Diez, Heiner Linke, and Dan~V. Nicolau.
\newblock Parallel computation with molecular-motor-propelled agents in nanofabricated networks.
\newblock {\em Proceedings of the National Academy of Sciences}, 113(10):2591--2596, 2016.

\bibitem{shiraishi2018speed}
Naoto Shiraishi, Ken Funo, and Keiji Saito.
\newblock Speed limit for classical stochastic processes.
\newblock {\em Physical review letters}, 121(7):070601, 2018.

\bibitem{vu2023topological}
Tan~Van Vu and Keiji Saito.
\newblock Topological speed limit.
\newblock {\em PHYSICAL REVIEW LETTERS}, 130(1), 2023.

\bibitem{barato2015thermodynamic}
Andre~C Barato and Udo Seifert.
\newblock Thermodynamic uncertainty relation for biomolecular processes.
\newblock {\em Physical review letters}, 114(15):158101, 2015.

\bibitem{horowitz2020thermodynamic}
Jordan~M Horowitz and Todd~R Gingrich.
\newblock Thermodynamic uncertainty relations constrain non-equilibrium fluctuations.
\newblock {\em Nature Physics}, 16(1):15--20, 2020.

\bibitem{van2022unified}
Tan Van~Vu, Yoshihiko Hasegawa, et~al.
\newblock Unified thermodynamic--kinetic uncertainty relation.
\newblock {\em Journal of Physics A: Mathematical and Theoretical}, 55(40):405004, 2022.

\bibitem{PhysRevLett.119.170601}
Todd~R. Gingrich and Jordan~M. Horowitz.
\newblock Fundamental bounds on first passage time fluctuations for currents.
\newblock {\em Phys. Rev. Lett.}, 119:170601, Oct 2017.

\bibitem{PhysRevLett.125.120604}
Gianmaria Falasco and Massimiliano Esposito.
\newblock Dissipation-time uncertainty relation.
\newblock {\em Phys. Rev. Lett.}, 125:120604, Sep 2020.

\bibitem{Manzano2024prx}
Gonzalo Manzano, G\"ulce Karde\ifmmode~\mbox{\c{s}}\else \c{s}\fi{}, \'Edgar Rold\'an, and David~H. Wolpert.
\newblock Thermodynamics of computations with absolute irreversibility, unidirectional transitions, and stochastic computation times.
\newblock {\em Phys. Rev. X}, 14:021026, May 2024.

\bibitem{yadav_circuits_2025}
Abhishek Yadav, Mahran Yousef, and David Wolpert.
\newblock Minimal thermodynamic cost of computing with circuits.
\newblock {\em arXiv preprint arXiv:2504.04031}, 2025.

\bibitem{yadav_rasp_2025}
Abhishek Yadav, Francesco Caravelli, and David Wolpert.
\newblock Mismatch cost of computing: from circuits to algorithms.
\newblock {\em arXiv preprint arXiv:2411.16088}, 2024.

\bibitem{yadav2024minimal_channels}
Abhishek Yadav and David Wolpert.
\newblock Minimal thermodynamic cost of communication.
\newblock {\em arXiv preprint arXiv:2410.14920}, 2024.

\bibitem{wolpert2019stochastic}
David~H Wolpert.
\newblock The stochastic thermodynamics of computation.
\newblock {\em Journal of Physics A: Mathematical and Theoretical}, 52(19):193001, 2019.
\newblock See arXiv:1905.05669 for updated version.

\bibitem{kolchinsky2020thermodynamic}
Artemy Kolchinsky and David~H Wolpert.
\newblock Thermodynamic costs of {T}uring machines.
\newblock {\em Physical Review Research}, 2(3):033312, 2020.

\bibitem{wolpert2020thermodynamics}
David~H Wolpert and Artemy Kolchinsky.
\newblock Thermodynamics of computing with circuits.
\newblock {\em New Journal of Physics}, 22(6):063047, 2020.

\bibitem{Ouldridge_2023}
Thomas~E Ouldridge and David~H Wolpert.
\newblock Thermodynamics of deterministic finite automata operating locally and periodically.
\newblock {\em New Journal of Physics}, 25(12):123013, dec 2023.

\bibitem{von1966theory}
John Von~Neumann, Arthur~Walter Burks, et~al.
\newblock Theory of self-reproducing automata.
\newblock 1966.

\bibitem{LANGTON1986120}
Christopher~G Langton.
\newblock Studying artificial life with cellular automata.
\newblock {\em Physica D: Nonlinear Phenomena}, 22(1):120--149, 1986.
\newblock Proceedings of the Fifth Annual International Conference.

\bibitem{johnston2022conway}
Nathaniel Johnston and Dave Greene.
\newblock {\em Conway's Game of Life: Mathematics and Construction}.
\newblock Nathaniel Johnston, 2022.

\bibitem{GREIF2024126}
Hajo Greif, Adam~P. Kubiak, and Pawe{\l} Stacewicz.
\newblock Selection, growth and form. turing's two biological paths towards intelligent machinery.
\newblock {\em Studies in History and Philosophy of Science}, 106:126--135, 2024.

\bibitem{wolfram1984cellular}
Stephen Wolfram.
\newblock Cellular automata as models of complexity.
\newblock {\em Nature}, 311(5985):419--424, 1984.

\bibitem{WOLFRAM19841}
Stephen Wolfram.
\newblock Universality and complexity in cellular automata.
\newblock {\em Physica D: Nonlinear Phenomena}, 10(1):1--35, 1984.

\bibitem{cook2004universality}
Matthew Cook et~al.
\newblock Universality in elementary cellular automata.
\newblock {\em Complex systems}, 15(1):1--40, 2004.

\bibitem{PROKOPENKO2019134}
Mikhail Prokopenko, Michael Harr{\'e}, Joseph Lizier, Fabio Boschetti, Pavlos Peppas, and Stuart Kauffman.
\newblock Self-referential basis of undecidable dynamics: From the liar paradox and the halting problem to the edge of chaos.
\newblock {\em Physics of Life Reviews}, 31:134--156, 2019.
\newblock Physics of Mind.

\bibitem{Prokopenko_2025}
Mikhail Prokopenko, Paul C~W Davies, Michael Harr{\'e}, Marcus~G Heisler, Zdenka Kuncic, Geraint~F Lewis, Ori Livson, Joseph~T Lizier, and Fernando~E Rosas.
\newblock Biological arrow of time: emergence of tangled information hierarchies and self-modelling dynamics.
\newblock {\em Journal of Physics: Complexity}, 6(1):015006, jan 2025.

\bibitem{schumann2024}
Andrew Schumann, Andrew Adamatzky, Jerzy Kr{\'o}l, and Eric Goles.
\newblock Fungi as turing automata with oracles.
\newblock {\em Royal Society Open Science}, 11(10):240768, 2024.

\bibitem{moore1991generalized}
Cristopher Moore.
\newblock Generalized shifts: unpredictability and undecidability in dynamical systems.
\newblock {\em Nonlinearity}, 4(2):199, 1991.

\bibitem{cardona2021constructing}
Robert Cardona, Eva Miranda, Daniel Peralta-Salas, and Francisco Presas.
\newblock Constructing turing complete euler flows in dimension 3.
\newblock {\em Proceedings of the National Academy of Sciences}, 118(19):e2026818118, 2021.

\bibitem{gonzalez2025topological}
{\'A}ngel Gonz{\'a}lez-Prieto, Eva Miranda, and Daniel Peralta-Salas.
\newblock Topological kleene field theories: A new model of computation, 2025.

\bibitem{KURKA1997203}
Petr K{\r u}rka.
\newblock On topological dynamics of turing machines.
\newblock {\em Theoretical Computer Science}, 174(1):203--216, 1997.

\bibitem{delvenne2005computational}
Jean-Charles Delvenne, Petr K{u}rka, and Vincent~D Blondel.
\newblock Computational universality in symbolic dynamical systems.
\newblock In {\em Machines, Computations, and Universality: 4th International Conference, MCU 2004, Saint Petersburg, Russia, September 21-24, 2004, Revised Selected Papers 4}, pages 104--115. Springer, 2005.

\bibitem{garland2011predicting}
Joshua Garland and Elizabeth Bradley.
\newblock Predicting computer performance dynamics.
\newblock In {\em International Symposium on Intelligent Data Analysis}, pages 173--184. Springer, 2011.

\bibitem{alexander2010}
Zachary Alexander, Todd Mytkowicz, Amer Diwan, and Elizabeth Bradley.
\newblock Measurement and dynamical analysis of computer performance data.
\newblock In Paul~R. Cohen, Niall~M. Adams, and Michael~R. Berthold, editors, {\em Advances in Intelligent Data Analysis IX}, pages 18--29, Berlin, Heidelberg, 2010. Springer Berlin Heidelberg.

\bibitem{piccinini2010computation}
Gualtiero Piccinini and Corey Maley.
\newblock Computation in physical systems.
\newblock 2010.

\bibitem{moore1990unpredictability}
Cristopher Moore.
\newblock Unpredictability and undecidability in dynamical systems.
\newblock {\em Physical Review Letters}, 64(20):2354, 1990.

\bibitem{lloyd1992any}
Seth Lloyd.
\newblock Any nonlinear gate, with linear gates, suffices for computation.
\newblock {\em Physics Letters A}, 167(3):255--260, 1992.

\bibitem{lloyd2006programming}
Seth Lloyd.
\newblock {\em Programming the universe: a quantum computer scientist takes on the cosmos}.
\newblock Vintage, 2006.

\bibitem{shiraishi2021undecidability}
Naoto Shiraishi and Keiji Matsumoto.
\newblock Undecidability in quantum thermalization.
\newblock {\em Nature communications}, 12(1):1--7, 2021.

\bibitem{cubitt2015undecidability}
Toby~S Cubitt, David Perez-Garcia, and Michael~M Wolf.
\newblock Undecidability of the spectral gap.
\newblock {\em Nature}, 528(7581):207--211, 2015.

\bibitem{pour1982noncomputability}
Marian~Boykan Pour-El and Ian Richards.
\newblock Noncomputability in models of physical phenomena.
\newblock {\em International Journal of Theoretical Physics}, 21(6):553--555, 1982.

\bibitem{davis1956note}
Martin Davis.
\newblock A note on universal turing machines.
\newblock {\em Automata studies}, 34:167--175, 1956.

\bibitem{sipser2012introduction}
Michael Sipser.
\newblock Introduction to the theory of computation. cengage learning.
\newblock {\em International edition}, 2012.

\bibitem{delvenne2009universal}
Jean-Charles Delvenne.
\newblock What is a universal computing machine?
\newblock {\em Applied Mathematics and Computation}, 215(4):1368--1374, 2009.

\bibitem{Fredkin1982}
Edward Fredkin and Tommaso Toffoli.
\newblock Conservative logic.
\newblock {\em International Journal of Theoretical Physics}, 21(3):219--253, Apr 1982.

\bibitem{chen2023rate}
Ho-Lin Chen, David Doty, Wyatt Reeves, and David Soloveichik.
\newblock Rate-independent computation in continuous chemical reaction networks.
\newblock {\em Journal of the ACM}, 70(3), May 2023.

\bibitem{doty2024}
Kim Calabrese and David Doty.
\newblock Rate-independent continuous inhibitory chemical reaction networks are turing-universal.
\newblock In Da-Jung Cho and Jongmin Kim, editors, {\em Unconventional Computation and Natural Computation}, pages 104--118, Cham, 2024. Springer Nature Switzerland.

\bibitem{minsky1967computation}
Marvin~Lee Minsky.
\newblock {\em Computation}.
\newblock Prentice-Hall Englewood Cliffs, 1967.

\bibitem{tao2019}
Terence Tao.
\newblock Searching for singularities in the navier--stokes equations.
\newblock {\em Nature Reviews Physics}, 1(7):418--419, 2019.

\bibitem{CARDONA2023109142}
Robert Cardona, Eva Miranda, Daniel Peralta-Salas, and Francisco Presas.
\newblock Universality of euler flows and flexibility of reeb embeddings.
\newblock {\em Advances in Mathematics}, 428:109142, 2023.

\bibitem{CARDONA202350}
Robert Cardona, Eva Miranda, and Daniel Peralta-Salas.
\newblock Computability and beltrami fields in euclidean space.
\newblock {\em Journal de Math{\'e}matiques Pures et Appliqu{\'e}es}, 169:50--81, 2023.

\bibitem{dyhr2025}
S{\o}ren Dyhr, {\'A}ngel Gonz{\'a}lez-Prieto, Eva Miranda, and Daniel Peralta-Salas.
\newblock Turing complete navier-stokes steady states via cosymplectic geometry, 2025.

\bibitem{González-Prieto_2025}
Ángel González-Prieto, Eva Miranda, and Daniel Peralta-Salas.
\newblock Universality in computable dynamical systems: old and new.
\newblock {\em Journal of Physics: Complexity}, 6(3):035014, sep 2025.

\bibitem{cardona2025towards}
Robert Cardona, Eva Miranda, and Daniel Peralta-Salas.
\newblock Towards a fluid computer.
\newblock {\em Foundations of Computational Mathematics}, pages 1--17, 2025.

\bibitem{Bausch2021}
Johannes Bausch, Toby~S. Cubitt, and James~D. Watson.
\newblock Uncomputability of phase diagrams.
\newblock {\em Nature Communications}, 12(1):452, Jan 2021.

\bibitem{Cubitt2015}
Toby~S. Cubitt, David Perez-Garcia, and Michael~M. Wolf.
\newblock Undecidability of the spectral gap.
\newblock {\em Nature}, 528(7581):207--211, Dec 2015.

\bibitem{compositionality:14134}
Tom{\'a}{\v s} Gonda, Tobias Reinhart, Sebastian Stengele, and Gemma~De les Coves.
\newblock A framework for universality in physics, computer science, and beyond.
\newblock {\em Compositionality}, Volume 6 (2024), Aug 2024.

\bibitem{purcell2024chaitin}
James Purcell, Zhi Li, and Toby Cubitt.
\newblock Chaitin phase transition, 2024.

\bibitem{SIEGELMANN1996461}
Hava~T. Siegelmann.
\newblock The simple dynamics of super turing theories.
\newblock {\em Theoretical Computer Science}, 168(2):461--472, 1996.

\bibitem{pour2017computability}
Marian~B Pour-El and J~Ian Richards.
\newblock {\em Computability in analysis and physics}, volume~1.
\newblock Cambridge University Press, 2017.

\bibitem{copeland2018church}
B~Jack Copeland and Oron Shagrir.
\newblock The church-turing thesis: logical limit or breachable barrier?
\newblock {\em Communications of the ACM}, 62(1):66--74, 2018.

\bibitem{kolchinsky2021}
Artemy Kolchinsky and David~H. Wolpert.
\newblock Work, entropy production, and thermodynamics of information under protocol constraints.
\newblock {\em Phys. Rev. X}, 11:041024, Nov 2021.

\bibitem{hartle2024work}
Harrison Hartle, David Wolpert, Andrew Stier, Christopher~P. Kempes, and Gonzalo Manzano.
\newblock Work extraction with feedback control using limited resources, 2024.

\bibitem{hartle2025distributed}
Harrison Hartle, Gonzalo Manzano, Jan Korbel, David Wolpert, Andrew Stier, Christopher~P. Kempes, and Gonzalo Manzano.
\newblock Distributed thermodynamic quenches, 2025.

\bibitem{kolchinsky_semantic_2018}
Artemy Kolchinsky and David~H. Wolpert.
\newblock Semantic information, autonomous agency and non-equilibrium statistical physics.
\newblock {\em Interface Focus}, 8(6):20180041, December 2018.

\bibitem{WolpertKolchinsky2016}
David~H. Wolpert and Artemy Kolchinsky.
\newblock Observers as systems that acquire information to stay out of equilibrium.
\newblock In {\em ``The Physics of the Observer'' Conference}, Banff, Canada, 2016.

\bibitem{karp1992line}
Richard~M Karp.
\newblock On-line algorithms versus off-line algorithms: How much.
\newblock {\em Algorithms, Software, Architecture: Information Processing}, 92:416, 1992.

\bibitem{Kosse2019}
Christin Kosse and Denis Burdakov.
\newblock Natural hypothalamic circuit dynamics underlying object memorization.
\newblock {\em Nature Communications}, 10(1):2505, Jun 2019.

\bibitem{bornschein2013v1}
J{\"o}rg Bornschein, Marc Henniges, and J{\"o}rg L{\"u}cke.
\newblock Are v1 simple cells optimized for visual occlusions? a comparative study.
\newblock {\em PLoS computational biology}, 9(6):e1003062, 2013.

\bibitem{li2025measuring}
Junang Li, Andrew~M Leifer, and David~H Wolpert.
\newblock Measuring amount of computation done by c. elegans using whole brain neural activity.
\newblock {\em arXiv preprint arXiv:2504.10300}, 2025.

\bibitem{mota2013sophistication}
Francisco Mota, Scott Aaronson, Lu{\'\i}s Antunes, and Andr{\'e} Souto.
\newblock Sophistication as randomness deficiency.
\newblock In {\em International Workshop on Descriptional Complexity of Formal Systems}, pages 172--181. Springer, 2013.

\bibitem{downey2019computability}
Rod Downey and Denis~R Hirschfeldt.
\newblock Computability and randomness.
\newblock {\em Notices of the American Mathematical Society}, 66(7):1001--1012, 2019.

\bibitem{ziv1978compression}
Jacob Ziv and Abraham Lempel.
\newblock Compression of individual sequences via variable-rate coding.
\newblock {\em IEEE transactions on Information Theory}, 24(5):530--536, 1978.

\bibitem{ZENIL201632}
Hector Zenil, Narsis~A. Kiani, and Jesper Tegn{\'e}r.
\newblock Methods of information theory and algorithmic complexity for network biology.
\newblock {\em Seminars in Cell \& Developmental Biology}, 51:32--43, 2016.
\newblock Information Theory in Systems Biology Xenopus as a model system for vertebrate development.

\bibitem{estevez2015lempel}
E~Estevez-Rams, R~Lora-Serrano, CAJ Nunes, and B~Arag{\'o}n-Fern{\'a}ndez.
\newblock Lempel-ziv complexity analysis of one dimensional cellular automata.
\newblock {\em Chaos: An Interdisciplinary Journal of Nonlinear Science}, 25(12), 2015.

\bibitem{haslinger2010computational}
Robert Haslinger, Kristina~Lisa Klinkner, and Cosma~Rohilla Shalizi.
\newblock The computational structure of spike trains.
\newblock {\em Neural computation}, 22(1):121--157, 2010.

\bibitem{cover2006elements}
Thomas Cover and Ray Thomas.
\newblock {\em Elements of information theory}.
\newblock John Wiley \& Sons, 2006.

\bibitem{li2025}
Junang Li, Andrew~M. Leifer, and David~H. Wolpert.
\newblock Measuring amount of computation done by c.elegans using whole brain neural activity, 2025.

\bibitem{takens2006detecting}
Floris Takens.
\newblock Detecting strange attractors in turbulence.
\newblock In {\em Dynamical Systems and Turbulence, Warwick 1980: proceedings of a symposium held at the University of Warwick 1979/80}, pages 366--381. Springer, 2006.

\bibitem{roscoe2010understanding}
Andrew~W Roscoe.
\newblock {\em Understanding concurrent systems}.
\newblock Springer Science \& Business Media, 2010.

\bibitem{reisig2016petri}
Wolfgang Reisig.
\newblock Petri nets.
\newblock In {\em Modeling in Systems Biology: The Petri Net Approach}, pages 37--56. Springer, 2016.

\end{thebibliography}

\end{document}